# Undergraduate Researcher Personas

Tra Huynh, trahuynh@ksu.edu, Kansas State University
Adrian M. Madsen, amadsen@aapt.org, American Association of Physics Teachers
Eleanor C. Sayre, esayre@ksu.edu, Kansas State University

**Abstract:** Personalized undergraduate research programs can help increase undergraduate students' participation in research. Personas of undergraduate researchers are a powerful means of encapsulating the richness of students' various goals, motivations, and experiences in research. Student personas can support the design of research programs with student-centered approaches that better engage undergraduate students across multiple disciplines. We used interview data and drew on Self-Determination Theory to identify students' various goals and motivations in order to build three undergraduate researcher personas in physics.

## Introduction

Authentic undergraduate research activities have a great impact on students' comprehensive development that neither lab courses nor other activities have been able to offer (Thiry et al., 2011; Hunter et al., 2007). Faculty-mentored, hands-on research programs offer a community of practice that supports science students' cognitive and personal growth, as well as the development of their professional identity (Irving & Sayre, 2015). Moreover, undergraduate research experiences also positively influence underrepresented students' retention, persistence, and pursuit of science career pathways (Barlow & Villarejo, 2004). Researchers have developed a huge body of work investigating undergraduate research experiences in STEM fields. In this work, we are seeking to solve the narrower problem of using student interviews to guide the design of undergraduate research programs.

Despite various attempts to include more students in undergraduate research (Sandnes et al. 2006), faculty still often report obstacles, including difficulties in pitching their research program to a wide variety of students. Although psychologists have shown the critical influence of motivation on students' attitudes and behaviors, little discussion has explicitly included the diversity of student motivations into designing research programs. As many of the tasks we want students to perform, including research practices, might not be inherently interesting to all students (Ryan & Deci, 2000), we need to account for the diversity of motivators when designing research programs. We argue that persona methodology can help with our educational context.

We repurposed existing interview data with undergraduate physics students to build personas of students participating in undergraduate research. Personas are life-like archetypes that stress users' diverse goals and motivations and embrace their corresponding needs and challenges. We use Self-Determination Theory to thoroughly describe the motivations that student personas have with regards to their research experiences. We claim that a set of memorable, sensible, and relatable personas can help facilitate student-centered discussions among faculty and departments with the goal of designing research programs that better fit students' goals and needs. For this design problem, we focus on building personas of students who are sufficiently motivated to engage in research, either intrinsically or extrinsically. We did not build personas of students who lack motivation to engage in research since it is not our goal to move students from unmotivated to motivated. Designing a complete research program with the constructed set of personas is also beyond the scope of this paper.

## Personas and Self-Determination Theory

The idea of personas originated from Alan Cooper's perspective that a design is meaningful only when people use it to achieve specific goals (Blomkvist, 2007). Personas are life-like models, whose characteristics are driven by the various goals and motivations of real or potential users. Personas help designers predict users' actions and behaviors, and thereby guide the design to assist users to accomplish their goals.

The persona approach is consistent with Self-Determination Theory (SDT) (Ryan & Deci, 2000), which emphasizes the critical role of motivation in energizing one's behavior and development. Although other classic perspectives view motivation as a continuous spectrum from low to high, SDT characterizes extrinsic and intrinsic motivations along axes of competence, autonomy, and relatedness. Your action is intrinsically motivated if you find the activity inherently enjoyable and satisfying. Conversely, your action is extrinsically motivated if you perform the action to gain separable outcomes, such as reaping instrumental rewards or avoiding sanctions. Each type of motivation yields different experiences and attitudes toward an activity. We drew on SDT to identify students' various goals and motivations and infer the corresponding experiences when building the personas.

Notably, personas are neither the descriptions of individuals nor average information of specific groups of users. Instead, personas are synthesized patterns of rich behaviors and motivations from an amalgamation of users (Madsen et al., 2014). Therefore, although the set of personas fully cover the data of real people, personal identities are highly protected compared to case studies. Personas also have some fictional details added, including name, picture, and background information, to make them more concrete and life-like.

Personas represent real users throughout the design process. Carefully crafted personas help designers put aside their biases and stereotypes of users and instead pay attention to whom the products are designed for and why or how those products will be used. By thinking about users first and foremost, designers can avoid urges to jump into fascinating design ideas that oftentimes are appealing to the designers but are not meaningful to the user (Blomkvist, 2007). Personas are powerful because they seem like real humans with realistic characteristics and stories; this evokes designers' empathy for real users and thereby, promotes user-centered design. Although the persona methodology has been mostly used in website design, much of the work of building personas has benefits beyond that field (Madsen et al., 2014; 2019). We aim this work toward building personas of undergraduate researchers, so that we gain a better understanding and representation of those researchers and subsequently support future design of student-centered research programs.

## Context and methodology

The data are drawn from semi-structured interviews with 2nd- and 3rd-year physics students at Kansas State University that was collected for a series of studies on students' identity formation, epistemological sophistication, and metacognition (Irving & Sayre, 2016; 2015). We conducted interviews with 21 students (18 male); 9 of them (8 male) joined research groups either before or during their interview sets. The gender ratio is typical in our department. The interview protocol explored students' interest and experience in physics, their perception of physicists, their self-perception and physics identity, and their professional career plans. We found this set of data well suited for our focus, and therefore we repurposed it to build student personas that address our design problem. We discuss some of the constraints from using this data set in later sections.

Our physics department offers undergraduate research opportunities in two forms: voluntary assistantships with or without stipends during the academic year and the NSF-funded summer Research Experience for Undergraduates program. Students make their own decisions to engage in research taken outside of class work; this makes their motivation towards research activity worthwhile for deeper investigation. Here, we purposefully built student personas from the 9 students with research experience. We argue that these students were sincerely motivated to involve themselves in research. Therefore, their motivations and experiences with research produce a reliable primitive set of personas.

Previously, the interview data had been phenomenographically (Marton, 1986) analyzed (Iving & Sayre, 2016; 2015); we drew on and extended the results of that analysis as we searched for information related to students' various research motivations and experiences. Building personas from phenomenographic study is a new approach; the details on the principles of this approach are discussed in our other work. After watching 5 interviews, we found emergent themes of college majors and minors; disciplinary experiences and motivations; physics identity; and career plans and awareness of other physics-major jobs. The whole set of interviews was repeatedly analyzed to explore the variations within each theme. For example, we found three distinct motivations for students engaged in research from the interview data, aligned with the autonomy, competence, and relatedness dimensions from SDT. In this context, autonomy concerns students' sense of volition and self-endorsement that stems from students' interests and integrated values. Competence refers to students feeling effective in interacting with the research and experiencing opportunities to exercise and express their capacity. Relatedness refers to students' feelings of connection with others in the research group and of belonging to the research community.

A set of personas is valid and useful when each student participant is primarily represented by one persona (with major goals, challenges, and core details) and secondarily represented by a few other personas (with minor motives and details). Personas are goal-driven models, so we prioritized motivations and goals when making them. We created a spreadsheet with a column of 3 different motives that potentially make up 3 personas. Each motivation set a tone that guided us to fill in each persona with characteristic variation from every phenomenographic theme. During this process, we made constant comparisons among the personas and the data to assure that these personas are distinct and meaningful, and collectively embody the phenomenographic results. After having sketches of the personas, we fleshed them out with names and short descriptive quotes and construed their potential actions and challenges. We then revised and validated the set of personas by matching back to the data before discussing the work with other researchers for peer review.

## Student personas

Although all students in our dataset expressed an inherent interest in physics, they had little idea about what research is like or what it entails before their involvement in research (Irving & Sayre, 2016), i.e. they were extrinsically motivated towards research activity. We constructed 3 personas – Maria, Ashley, Louis (Table 1) - whose motivations to do research are at various levels of autonomy, relatedness, and competence.

Table 1: Undergraduate researcher personas

| Personas | Description | Key quotes | Challenges |
|---|---|---|---|
| Maria | Has many academic and non-academic interests. Open to many future career options. Thinks that research is somewhat important to scientists. Engages in research for experiences of self-competence and intellectual exploration. | "I want to see what it's like." | Self-commitment and time commitment |
| Ashley | Ambivalent physics background; does not commit to physics until having a great physics class or mentors. Career plan undetermined and open, but has some role models. Thinks that field commitment to the field is important to scientists. Joins research group in order to work with people she likes. | "I want to work with these people." | Struggles with self-efficacy and feeling of belonging |
| Louis | An aspiring scientist who is determined on Research – Grad School – Professional career path. Thinks that research is significant to scientists. Engages in research to learn skills, seek for research interests, and get competent for grad school. | "I want to go to grad school." | Frustration with research activity or disappointment with self-efficacy |

Of all the personas, Maria has the most diverse interests, which includes physics and physics research. Maria-type students are likely to have double majors or multiple minors, one of which is physics. She thinks physics research is somewhat important to physicists; therefore, she wants to try it out. Multiple interests also mean that she is open to many options and may have trouble deciding which option to try. This could be an obstacle to her committing to a research experience, regardless of her interest in physics. Maria is concerned with how research activity fits her broader interests and values, and she spends time weighing physics research with other options, ranging from research in other disciplines to non-academic activities. She is most attracted to research that is fun, where she can learn interesting things, and which does not require too much commitment.

Ashley stands out as a persona with a strong sense of relatedness, who is extrinsically motivated by the social influence of research activity. Ashley-type students might have had unfavorable experiences in other departments, or they might not have committed to physics until meeting a welcoming mentor or taking a great physics class. She participates in research for the fulfillment of relatedness needs, on the scope of either working with a direct mentor or with the physics community. Ashley tends to desire approval from her advisor and research colleagues as rewards and has not integrated extrinsic motivation yet. Therefore, Ashley might later encounter competence demotivators, which likely diminishes feelings of relatedness, i.e. feeling left behind, feeling incompetent at doing research.

Louis is pretty determined on the path of Research – Grad School – Professional Career; he sees research as a necessary step to pursue post-graduate studies. Louis-like students do not necessarily have a specific research interest yet, but they are not hesitant to consider themselves as aspiring physicists, to approach research mentors, and engage in research early in their undergraduate program.

Ideally, the set of personas fully covers the aspects of students' experiences that are important to the design problem. Due to the space constraint, we only present here one brief excerpt to illustrate the persona of Maria. For example, a student named Ed (pseudonym) is slightly represented by Louis for planning to become an academic professional, but he is dominantly represented by Maria for his struggles with choosing between interests in physics and chemistry, as well as allocating time for another non-academic activity as well as for language courses. "I couldn't pick between either of [physics and chemistry]. They're both a lot of fun… Well… there is a possibility that A, I just won't ever decide. I'll be dabbling in everything forever. Or B, after trying different things, I'll find one that I specifically enjoy more than others."

## Discussion

We generated three personas of undergraduate researchers driven by various goals and motivations, which are differentiated by SDT. Personas are powerful because they encapsulate students' rich and relevant information in sensible and memorable forms. Using such a set of personas would not only help researchers, faculty, and departments to sympathize with and understand the richness of student variation but would also create a space for effective communication among them without revealing students' identity. These plausible and relatable personas can help to bolster student-centered research design as well as optimize research program designs that fit real students as opposed to our biased perceptions of students. When we understand students' motivations as represented through personas, we can properly predict their experience in undergraduate research, and design motivators and pedagogical advising practices that increase students' individual interest, satisfaction, and retention in research. In other words, personas can point the way to a successful pitch for research programs to diverse students while saving faculty time and effort.

For example, in our context, Louis is excited for research experience regardless of being paid or not. Louis-type students may just need information such as faculty project details and contacts, and faculty do not need to make an extended effort to include them. For Maria-type students, however, it is important to advertise the research in such a way that highlights its interesting aspects, promotes tips for self and time management, and encourages her to check it out. Ashley needs close contact and collaboration with research peers and seniors; their faculty mentor should provide them with supportive collaboration.

Notably, the frequency of persona appearance in data is not the most important factor to validate personas. Rather, it is how personas are distinct from each other and collectively expand the data set. In our data, we find more Louis-like students than Maria and Ashley-like ones. However, we predict that we would find more Maria and Ashley-like students if we expanded the sample data. Also, the interview protocol on this data did not purposefully explore students' family or financial backgrounds, and we expect these aspects might play a role in additional personas. For example, an additional persona might represent non-traditional students – who are age 25 or older, and who prioritize the need to earn money when making decisions about their academic activities (Becker, et al., 2003). This persona might be interested in physics research as providing part-time jobs or better future jobs, but they might also weigh the value of research opportunities against other internship opportunities as well as against non-academic jobs.

We also emphasize that personas are context and design-problem sensitive. We recognize that our data, even though it is characteristic for our institution, is limited and not representative for other contexts. In other words, while our persona set is suited to serve our institutional design problem, it may not be fully applicable and helpful in other contexts. We encourage departments and researchers to carefully justify any personas according to their targeted data. In addition, we advocate for the incorporation of personas methodology in educational research contexts for the advantage of strongly protecting participants' identity when research is presented to and discussed within departments.

**Acknowledgments**
We are greatly indebted to the researchers that conducted this interview data and the copy editor *Jeremy Smith*. This work is based upon work supported by Grant 1726479 and 1726113 and Kansas State University Physics Department.